\documentclass[a4paper,11pt]{article}
\usepackage{jinstpub} % for details on the use of the package, please see the JINST-author-manual
\usepackage{lineno}
\usepackage[UKenglish]{babel}
\usepackage{amsmath}
\usepackage{wasysym}[nointegrals]
\usepackage{siunitx}
\usepackage{xcolor}
\usepackage{hyperref}
\usepackage[labelformat=simple]{subcaption}
\usepackage{tikz}
\usepackage{textgreek}
\usepackage{pgfplots}
 \pgfplotsset{compat=1.16}
 \pgfdeclarelayer{background}
 \pgfdeclarelayer{foreground}
 \pgfsetlayers{background,main,foreground}
 \usepgfplotslibrary{fillbetween}
 \usepgfplotslibrary{dateplot}

\title{Characterization of large diameter ultra-thin vacuum windows for soft X-ray applications}

\author[a]{K. Desch,}
\author[b]{E. Ferrer-Ribas,}
\author[c]{F.J. Iguaz,}
\author[a]{J. von Oy,}
\author[d]{A. Quintana,}
\author[a,1]{T. Schiffer\note{Corresponding author.}}
\affiliation[a]{Physikalisches Institut,\\University of Bonn, 53115 Bonn, Germany}
\affiliation[b]{IRFU, CEA,\\Université Paris-Saclay, 91191 Gif-sur-Yvette, France}
\affiliation[c]{SOLEIL Synchrotron,\\L’Orme des Merisiers, Départementale 128, 91190 Saint-Aubin, France}
\affiliation[d]{Centro de Astropartículas y Física de Altas Energías (CAPA),\\University de Zaragoza, 50009 — Zaragoza, Spain}

\emailAdd{schiffer@physik.uni-bonn.de}

\abstract{We present novel, ultra-thin, large-diameter silicon nitride windows for various soft X-ray applications. Together with the company NORCADA, we developed windows with 200\,nm and 300\,nm thickness withstanding pressure differences above 1\,bar. The windows have an open diameter of 14\,mm. They were intensively vacuum- and overpressure-tested, showing very good results. At a measurement campaign at the synchrotron radiation source SOLEIL in France, the transparency of the windows was measured over a range from 50\,eV to 15\,keV, giving results comparable with the expected transparencies.}


\begin{document}
\maketitle
\flushbottom
\section{Introduction}
Soft X-rays might be a key to find new physical phenomena like solar axions~\cite{abeln2021conceptual, redondo2013solar}. They are also widely used for different imaging techniques at synchrotron radiation facilities~\cite{hitchcock2015soft, van2014x, collins2022resonant, el2023experimental}. Typically, silicon pixel detectors are used for such measurements~\cite{baruffaldi2025single}; however, they suffer from typically low quantum efficiencies, especially at the lower energy threshold. To circumvent this,  MicroMegas-like detectors ~\cite{castcollaboration2024new, krieger2017gridpix, desch2026construction} can be used, since they can achieve a quantum efficiency close to 100\,\% under optimized conditions. These detectors are filled with a gas and have to be coupled to a vacuum system. Therefore, a barrier, transparent for soft X-rays, separating the gas from the vacuum is necessary. Typical windows for such applications are Beryllium or Carbon fiber windows~\cite{huebner2015high, artyukov2004x} which typically have cutoffs below $3\,\text{keV}$. To reach even lower energies, so far mostly thin foils of polypropylene (PP) or Mylar\textsuperscript{\tiny\textregistered} were used. However, they need to have a certain thickness $\mathcal{O}(\text{\textmu m})$, lowering their transparency, to reach a reasonable vacuum tightness. Especially at larger diameters ($\text{\diameter}\geq 10\,\text{mm}$), windows made from polymers tend to have high leak rates. Compared to those classic windows, as shown in figure~\ref{SiN:Nominal_window}, ultra-thin silicon nitride ($\text{SiN}_\text{x}$) membranes exhibit a good transparency in the regime below $2\,\text{keV}$. In the following a characterization of such windows with a diameter of $14\,\text{mm}$ will be given. 

\section{Ultra-thin silicon nitride windows}\label{sec:sin_windows}
 To reach the needs of having a good transparency for soft X-rays and a high strength, while being vacuum-tight, silicon nitride ($\text{SiN}_\text{x}$, where $0 < x \leq 1.33$) thin films are a potent solution~\cite{kaloyeros2020silicon,lefevre1991thin,torma2013ultra,torma2014performance}, being on the one hand vacuum tight while on the other hand having a good mechanical stability in very thin $\mathcal{O}(100\,\text{nm})$ configurations. Albeit the properties of those membranes strongly depend on the way of production. For example, the density can vary over a range of $2.0\,\text{g}/\text{cm}^3$ to $3.44\,\text{g}/\text{cm}^3$~\cite{kaloyeros2020silicon}, having a direct impact on the transmission of the X-rays.

\begin{figure}[htb] 
\centering 
\begin{tikzpicture} 
    \begin{axis}[
                grid = major,
                xlabel={Energy [eV]},
                ylabel=Transparency,
				xmin=10, xmax=6000,
                xtick={1000,2000,...,6000},
				ymin = 0, ymax = 1.1,
				scaled y ticks=false,
				scaled x ticks=false,
				width=0.8\textwidth,
				height=0.5\textwidth,
				legend style={at={(0.97,0.03)},anchor=south east,cells={align=left}}
                ]
        \addplot[black, line width=0.5pt] table[skip first n=1, x index=0, y expr = \thisrowno{1}*\thisrowno{2}, col sep=space] {data/SiN_window_trans.csv};
        \addlegendentry{$200\,\text{nm}$ $\text{SiN}_\text{x}$ + $20\,\text{nm}$ Al}
		\addplot[blue, line width=0.5pt] table[skip first n=1, x index=0, y expr = \thisrowno{1}*\thisrowno{3}, col sep=space] {data/SiN_window_trans.csv};
		\addlegendentry{$300\,\text{nm}$ $\text{SiN}_\text{x}$ + $20\,\text{nm}$ Al}
		\addplot[red, line width=0.5pt] table[skip first n=1, x index=0, y expr = \thisrowno{1}*\thisrowno{6}, col sep=space] {data/SiN_window_trans.csv};
		\addlegendentry{$2\,\text{\textmu m}$ Mylar\textsuperscript{\tiny\textregistered} + $20\,\text{nm}$ Al}
		\addplot[green, line width=0.5pt] table[x index=0, y expr = \thisrowno{1}, col sep=space] {data/Be_25mu.txt};
		\addlegendentry{$25\,\text{\textmu m}$ Be}
	\end{axis}
\end{tikzpicture}  
\caption{Expected transparency of the ultra-thin silicon nitride windows ($\text{Si}\text{N}_{1.33}$) with a material density of $\rho = 3.44\,\text{g}\:\!\text{cm}^{-3}$. For comparison also the transmissions of a Mylar\textsuperscript{\tiny\textregistered} and a beryllium window are shown. It can be clearly seen that in the energy range below $2\,\text{keV}$ the $\text{SiN}_\text{1.33}$ windows perform much better than the Mylar\textsuperscript{\tiny\textregistered} or beryllium windows. Above $6\,\text{keV}$ the transparency does not change.Transparency data from~\cite{henke}.} 
\label{SiN:Nominal_window} 
\end{figure}
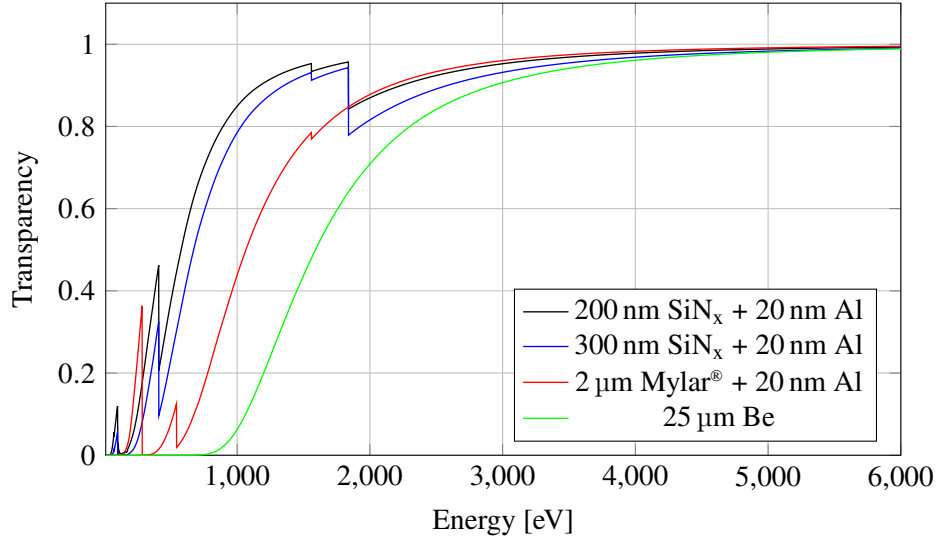

To find the best configuration, first, a simulation with COMSOL and SOLIDWORKS was performed. The design here was focused on getting the largest open area in the strong-back with the thinnest possible membrane. 
After this, a variety of samples were produced and tested. The findings of these tests were then fed back into the simulation, finally resulting in a design with $300\,\text{nm}$ thick membranes and a diameter of $14\,\text{mm}$, coated with a thin ($20\,\text{nm}$) aluminium layer. The testing included overpressure tests to ensure the quality of the window (section~\ref{ssec:windowdesign}) as well as helium leak tests to specify the vacuum tightness (section~\ref{ssec:windowvac}), resulting in a window design with an open area of 83.8\,\%. 

In this development process, $200\,\text{nm}$ windows were tested as well; however, only a single window could be found to survive multi-cycle overpressure tests to $1.2\,\text{bar}$. The expected transparencies for the two types of windows compared to the standard Mylar\textsuperscript{\tiny\textregistered} or beryllium windows are shown in figure~\ref{SiN:Nominal_window}.The performance of the $\text{SiN}_\text{x}$-windows decreases, as expected, with the window thickness.

\subsection{Design of the windows}\label{ssec:windowdesign}
\begin{figure}[htb]
\begin{subfigure}{0.5\linewidth}%1
  \centering
	\begin{tikzpicture}
    \node[anchor=south west,inner sep=0] (Bild) at (0,0) {\includegraphics[width=0.62\textwidth]{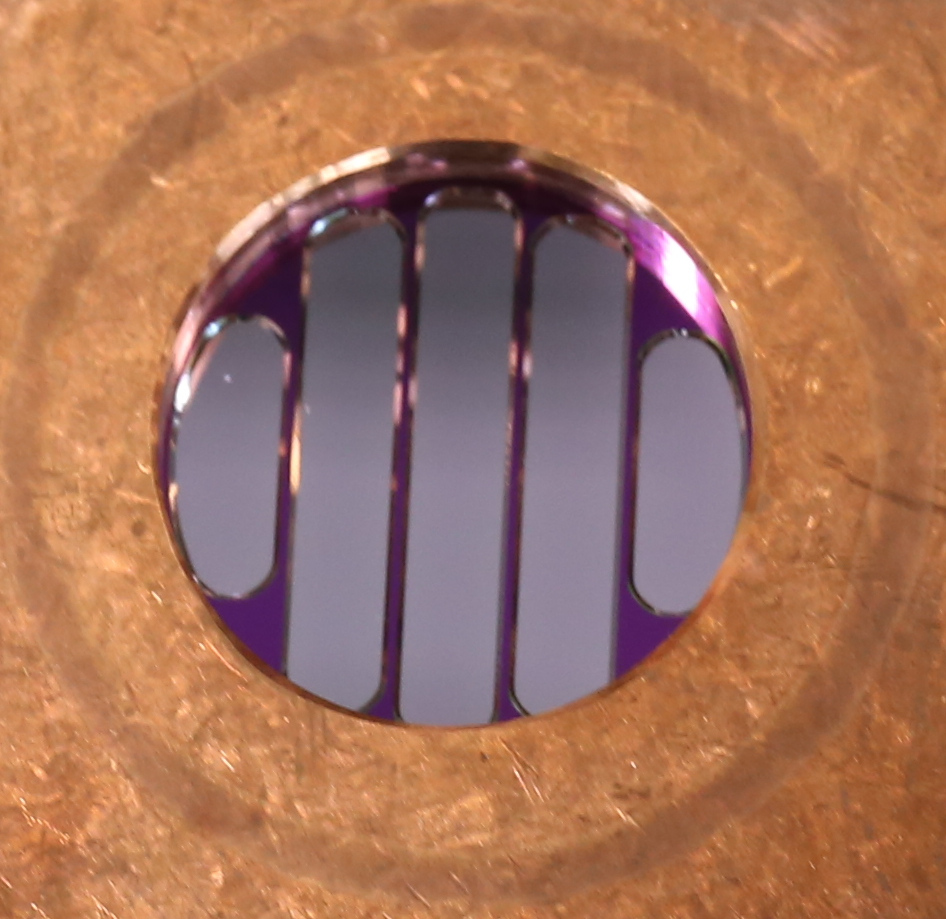}};
\end{tikzpicture}
\label{fig:strongback_a}
\caption{}
\end{subfigure}
\begin{subfigure}{0.5\linewidth}%1
  \centering
  \begin{tikzpicture}[font=\footnotesize]
    \node[anchor=south west,inner sep=0] (Bild) at (0,0) {\includegraphics[width=0.6\linewidth]{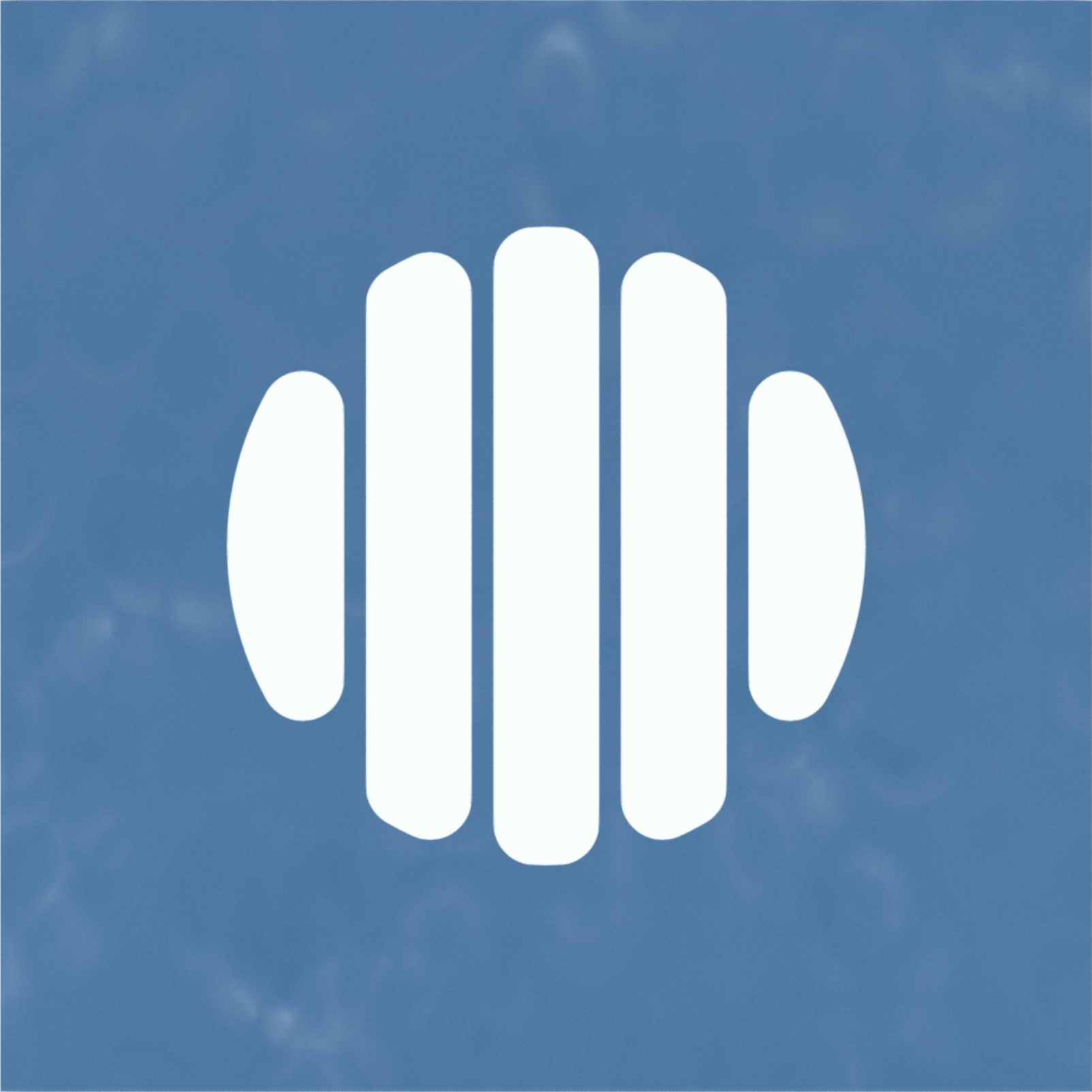}};
    \begin{scope}[x=(Bild.south east),y=(Bild.north west)]
      \node [right, align=left] (cat) at (1.05,0.5){{$24\,\text{mm}$}};
      \draw [|-|, thick, black](1.05,0) -- (1.05,1);
    \end{scope}
  \end{tikzpicture}
  \caption{}
  \label{fig:layout_strongback}
    \end{subfigure}
    \caption{a) Picture of low-pressure/vacuum side of the window (\text{\diameter}=14\,mm) with the strongback structure mounted in a copper holder. b) Layout of the strongback of the window.}
    \label{fig:strongback}
\end{figure}
The windows were designed as a combination of a silicon nitride membrane on a silicon strong back structure and produced in a proprietary process by the company Norcada~\cite{gridpix-norcada}. The $\text{SiN}_\text{x}$ membranes were tuned for quality, pressure robustness, transmission requirements at the transmission regimes required, first by choosing the right (proprietary) composition for the membrane layer, and second by ensuring the device layers were atomically bonded at every intersection. The membranes were formed using a non-conventional Silicon etch process to create the cavities without any sharp corners, and with support ribs optimized for the mechanical robustness required. The features were all defined using a UV lithography process, followed by etching per the above. Then the aluminium layer is deposited using a DC magnetron sputtering system, followed by a proprietary densification process. The current design is shown in figure~\ref{fig:strongback}. To make the windows light-tight and electrically conductive, the high-pressure side is coated with $20\,\text{nm}$ of aluminium.  
To check the reliability of the windows, a testing scheme with five to six pressure cycles from atmospheric pressure to $1.5\,\text{bar}$ ($1.2\,\text{bar}$) overpressure and back was performed for all $300\,\text{nm}$ ($200\,\text{nm}$) windows. During these tests, one cycle was always performed to stay at $1.5\,\text{bar}$ over several hours (for more detail see~\cite{SchifferPhD}).

\subsection{Vacuum testing}\label{ssec:windowvac}
\begin{figure}[htb]
\centering
	\begin{tikzpicture}
    \node[anchor=south west,inner sep=0] (Bild) at (0,0) {\includegraphics[width=0.3\textwidth]{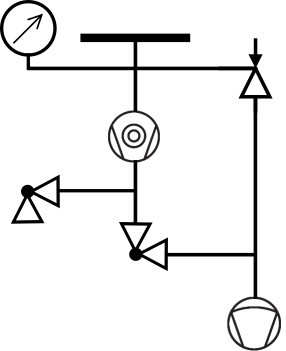}};
    \begin{scope}[x=(Bild.south east),y=(Bild.north west)]
    \node[anchor=west] at (0.55,0.6) {TMP};
    \node[anchor=west] at (1,0.075) {Backing pump};
    \node[anchor=west] at (0.97,0.8) {Needle valve};
    \node[anchor=east] at (0,0.91) {Gauge};
    \node[anchor=west] at (0.3,0.95) {Window};
    \node[anchor=east, align=left] at (0,0.45) {Connection to\\ leak tester};
    \end{scope}
\end{tikzpicture}
\caption{Schematic of the vacuum test stand for the ultra thin windows.}
\label{fig:test_stand}
\end{figure}
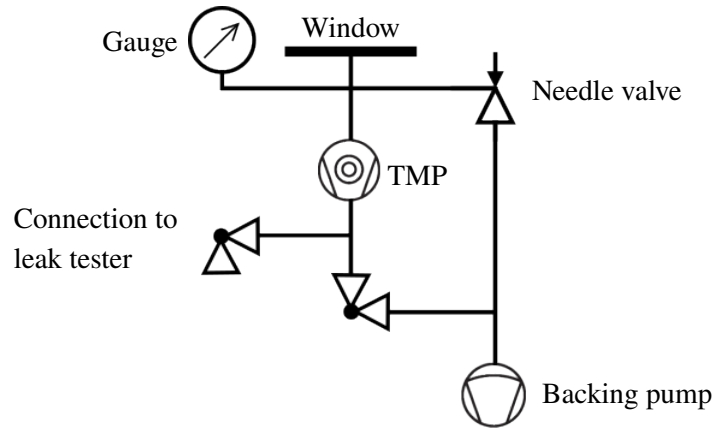
After the overpressure tests, the windows were mounted on a vacuum test bench to be helium leak tested. As shown in figure~\ref{fig:test_stand}, the test bench consists of a turbo-molecular pump (TMP) connected to a flange to mount the windows. The backing pump has an additional bypass with a needle valve to allow for a slow reduction of the pressure. All windows were successfully tested until the detection limit of the leak tester was reached. For one window with a $300\,\text{nm}$ membrane, a limit of better than $3.0\times10^{-9}\,\text{mbar\,l\,s}^{-1}$ was reached. Thus, the windows can be considered leak-tight. This was also shown by other studies~\cite{torma2013ultra, torma2014performance}. For the window with a $200\,\text{nm}$ membrane a leak rate below $8.0\times10^{-9}\,\text{mbar\,l\,s}^{-1}$ was measured. Hence, the leak tightness does not seem to decrease significantly with the thickness.

\section{Transparency measurements}\label{sec:sin_soleil}
To estimate the detection efficiency of X-rays in a detector utilizing these windows, their transparency has to be known. Theoretical calculations depend heavily on the ratios and thicknesses, which are not known precisely. Therefore, transparency measurements have been performed at the synchrotron radiation source SOLEIL~\cite{soleil}. Its METROLOGIE beamline~\cite{metro_bmln} supplies X-rays with energies between $30\,\text{eV}$ and $38\,\text{keV}$ which exceeds the desired energy range.

\subsection{The METROLOGIE beamline at SOLEIL}\label{ssec:Soleil}
The METROLOGIE beamline is one of 27 synchrotron radiation beamlines at SOLEIL, of which all have different purposes and energy ranges. The X-rays produced there originate as bending magnet radiation from electrons in the storage ring. These electrons are first accelerated to $100\,\text{MeV}$ in a linear accelerator (LINAC) and then further accelerated to $2.75\,\text{GeV}$ in a booster synchrotron. They are then injected into the final storage ring every time the beam current drops below a threshold in a top-up mode. The maximum beam current reached is $450\,\text{mA}$. This top-up mode is translated into the X-ray current in the METROLOGIE beamline and results in varying intensities of signal.\\
The beamline is split into two branches, the soft and the hard X-ray branch. Due to the broadness of the energy spectrum of the X-rays produced as bending magnet radiation, filter mechanisms are in place for the two branches. In the soft X-ray branch~\cite{idir2006technical} energies from $30\,\text{eV}$ to $2\,\text{keV}$ with a line width smaller than $1\,\text{eV}$~\cite{Khubbutdinov_2022, peatman2018gratings} can be selected using a gratings monochromator. The hard X-ray branch provides energies from $3\,\text{keV}$ to $38\,\text{keV}$ by a double crystal monochromator. The line width in this branch depends on the energy of the X-rays and can be described with $\Delta E/E = 10^{-4}$~\cite{willmott2019introduction}. For the hard X-ray branch, the alignment of the crystals also has to be slightly detuned to reduce higher harmonics negatively influencing the results~\cite{menesguen2011characterization}. Both beamlines supply beam spots on target smaller than $0.5\,\text{mm}$, thus sufficiently small to only test the membranes.

\subsection{Setup and data taking}\label{ssec:Datataking}
\begin{figure}[htb]
\centering
\begin{subfigure}[t]{0.45\textwidth}
\centering
	\begin{tikzpicture}
    \node[anchor=south west,inner sep=0] (Bild) at (0,0) {\includegraphics[width=0.9\textwidth]{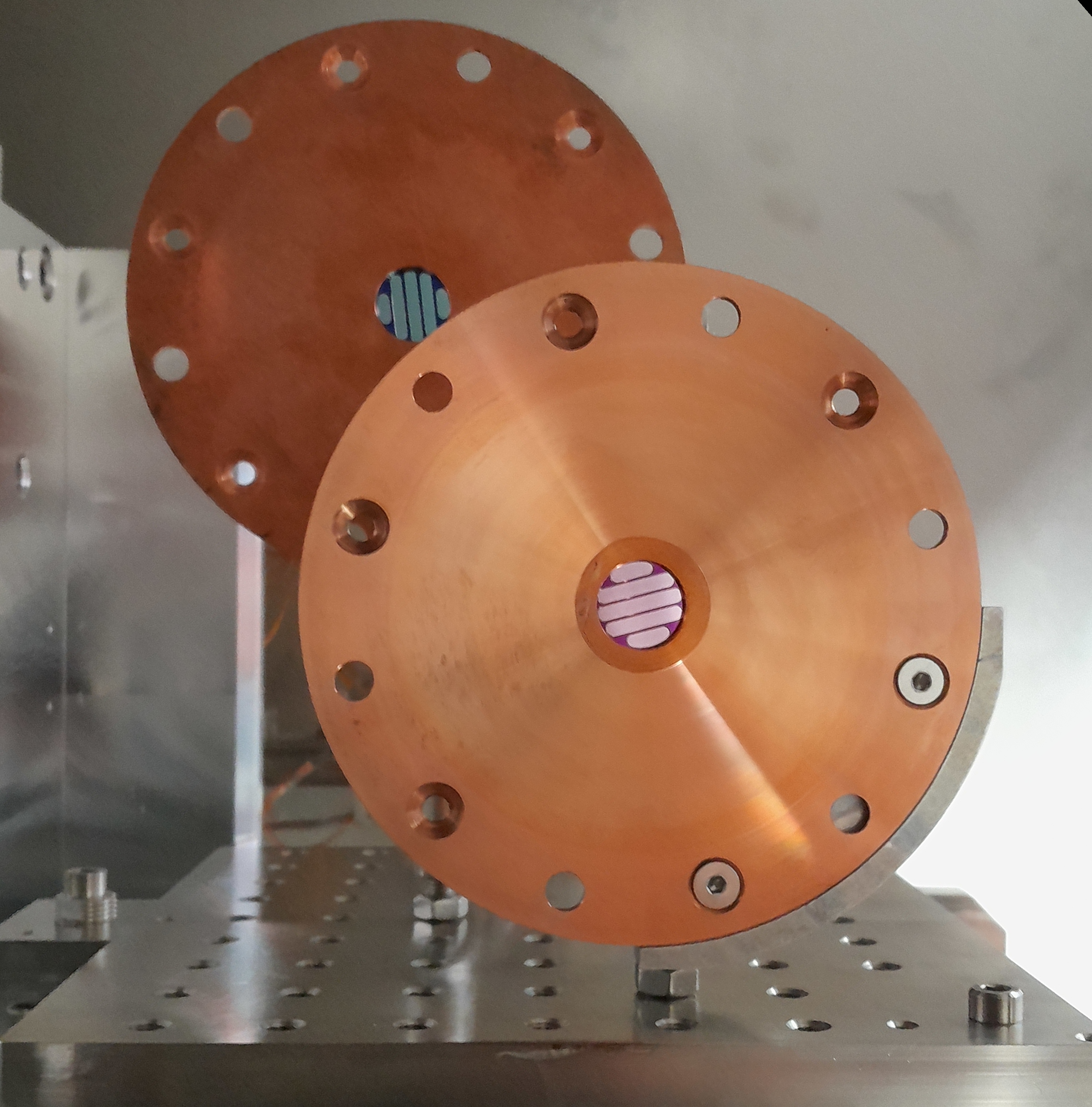}};
\end{tikzpicture}
\caption{}
  \label{fig:windows_hard}
\end{subfigure}%
\begin{subfigure}[t]{0.45\textwidth}
\begin{tikzpicture}
    \node[anchor=south west,inner sep=0, opacity=0.75] (Bild) at (0,0) {\includegraphics[width=0.9\textwidth]{figs/windows.jpg}};
    \begin{scope}[x=(Bild.south east),y=(Bild.north west)]
    \fill [blue!50] (0.3,0.2) circle (0.25cm);
    \node [] (1) at (0.3,0.2){1/4};
    \fill [blue!50] (0.59,0.45) circle (0.25cm);
    \node [] (2) at (0.59,0.45){2};
    \fill [blue!50] (0.37,0.72) circle (0.25cm);
    \node [] (3) at (0.37,0.72){3};
    \draw [-stealth, thick, black](1.north east) -- (2.south west);
    \draw [-stealth, thick, black](2.north west) -- (3.south east);
    \draw [-stealth, thick, black](3.south) -- (1.north);
    \end{scope}
\end{tikzpicture}
\caption{}
  \label{fig:windows_path}
\end{subfigure}
\caption{(a) Picture of the windows mounted in the vacuum tank of the hard X-ray branch. (b) Indicated positions for one measurement cycle 1-2-3-4, here 1 and 4 are calibration measurements to determine the total flux.}
\end{figure}

\begin{figure}[htb] 
\centering 
\begin{tikzpicture} 
    \begin{axis}[name=axis1,
                grid = major,
		        xlabel={Vertical position in [mm]},
                ylabel={Current [A]},
				xmin= -24, xmax = -7,
				scaled y ticks=false,
				width=0.8\linewidth,
				height=0.5\linewidth,
				legend style={at={(0.97,0.97)},anchor=north east,cells={align=left}}
                ]
        \addplot[black, line width=0.5pt, mark=none, mark size=1pt] table[x index=0,y index=1,col sep=tab] {data/pos_window_soleil.txt};
    \end{axis}
\end{tikzpicture}		
\caption{Position measurement of one window. The vertical direction is scanned to find the optimal position of the window membrane (center of one of the five flat tops). The four dips visible in the scan are originating from the four ribs of the strongback.}
\label{SiN:Pos_meas}
\end{figure}
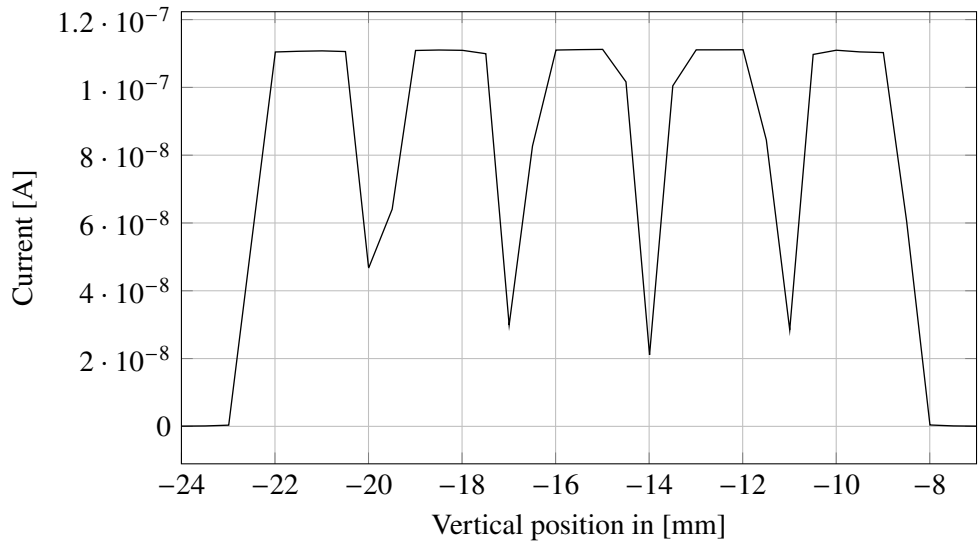
The soft and the hard X-ray branches both include a vacuum chamber where the windows were mounted in window holders on a movable platform as shown in figure~\ref{fig:windows_hard}. This allows for multiple windows to be tested at the same time, as well as measurements of the nominal flux without a window for each X-ray energy. Preceding the transparency measurements, each of the four windows had to undergo a spatial scan to align the center of the beam to an area of the window where the membrane is not obstructed by the strongback. This is exemplarily shown in figure~\ref{SiN:Pos_meas}. \\

\begin{figure}[htb] 
	\centering 
	\begin{tikzpicture} 
		\begin{axis}[name=axis1,
			grid = major,
			xlabel={Time [s]},
			ylabel={Current [A]},
			xmin= 0, xmax = 60,
			scaled y ticks=false,
			width=0.8\textwidth,
			height=0.5\textwidth,
			legend style={at={(0.97,0.97)},anchor=north east,cells={align=left}}
			]
			\addplot[black, line width=0.5pt, mark=none, mark size=1pt] table[x index=0,y index=1,col sep=comma] {data/Scan_700eV_200nm_edit.csv};
			\addlegendentry{Current}
			\addplot[ samples=5, smooth, blue] coordinates {(0,3.7814e-9)(60,3.7814e-9)};
			\addlegendentry{Mean}
			\addplot[ samples=5, smooth, draw=none, no markers, name path = std1] coordinates {(0,3.7868002e-9)(60,3.7868002e-9)};
			\addplot[ samples=5, smooth, draw=none, no markers, name path = std2] coordinates {(0,3.7759998e-9)(60,3.7759998e-9)};
			\addplot [fill=blue, opacity=0.2]fill between[of=std1 and std2];
		\end{axis}
	\end{tikzpicture}		
	\caption{Example of a measurement taken at the soft X-ray branch at an energy of~$700\,\text{eV}$. The mean ($\overline{x}_\text{diode, 700\,eV}=3.7814\times10^{-9}\,\text{A}$) and the standard deviation ($s_\text{diode, 700\,eV}=5.4002\times10^{-12}\,\text{A}$) used for the transparency calculations are indicated.}
	\label{SiN:Data}
\end{figure}
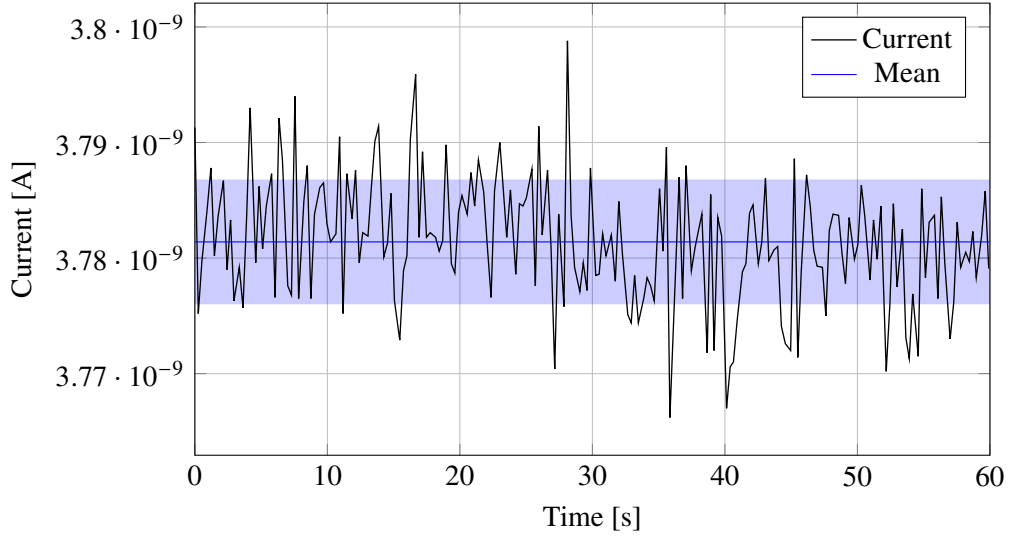

To measure the signal strength after the windows, a photodiode is mounted in the vacuum chamber (\texttt{IRD AXUV100}~\cite{IRD_Photodiode}). After setting the X-ray energy, each window's transparency was measured for one minute, including two calibration measurements in the beginning and at the end of each run. Such a measurement cycle is indicated in figure~\ref{fig:windows_path}. The transparency is then determined by comparing the measured photodiode current from each window with with the currents from the calibration measurements. An example of such a measurement can be seen in figure~\ref{SiN:Data}, with the mean value and its standard deviation indicated. At the soft X-ray branch, 32 different energies ranging from $50\,\text{eV}$ to $2\,\text{keV}$ were tested. Here, different settings of the monochromator grating, the filter, and the low-order sorter could be chosen to optimize the line width and flux of the X-ray beam. At the hard X-ray branch measurements at 15 energies ranging from $3\,\text{keV}$ to $15\,\text{keV}$ were taken, leaving a gap of $1\,\text{keV}$ between the branches. 

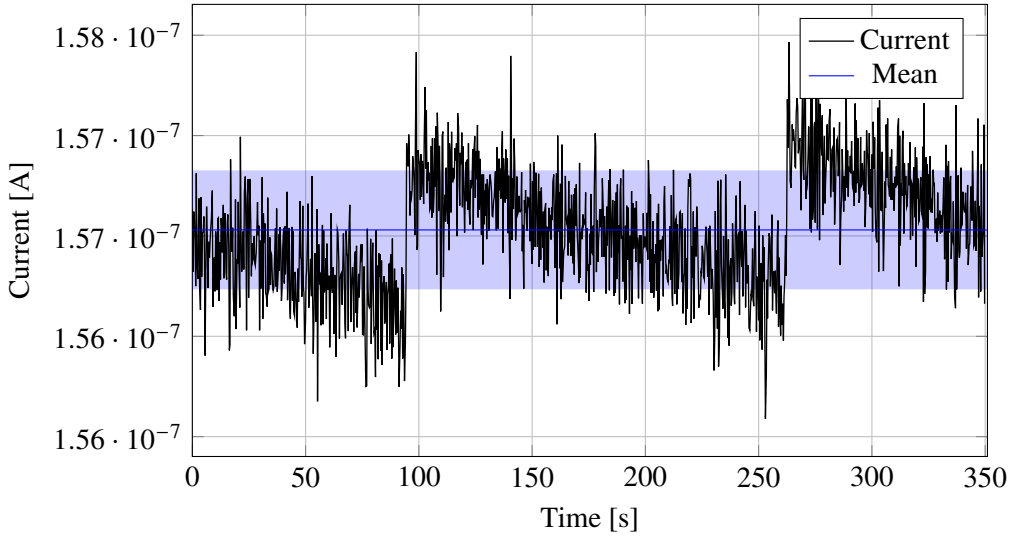
\begin{figure}[htb] 
\centering 
\begin{tikzpicture} 
    \begin{axis}[name=axis1,
                grid = major,
				xlabel={Time [s]},
                ylabel={Current [A]},
				xmin= 0, xmax = 351,
				scaled y ticks=false,
				width=0.8\linewidth,
				height=0.5\linewidth,
				legend style={at={(0.97,0.97)},anchor=north east,cells={align=left}}
                ]
        \addplot[black, line width=0.5pt, mark=none, mark size=1pt] table[x index=0,y index=1,col sep=comma] {data/Long_Time_Scan_350s_1.1KeV_edit.csv};
		\addlegendentry{Current}
		\addplot[ samples=5, smooth, blue] coordinates {(0,1.5653e-7)(351,1.5653e-7)};
		\addlegendentry{Mean}
		\addplot[ samples=5, smooth, draw=none, no markers, name path = std1] coordinates {(0,1.56234166e-7)(351,1.56234166e-7)};
		\addplot[ samples=5, smooth, draw=none, no markers, name path = std2] coordinates {(0,1.56826575e-7)(351,1.56826575e-7)};
		\addplot [fill=blue, opacity=0.2]fill between[of=std1 and std2];
    \end{axis}
\end{tikzpicture}		
\caption{Calibration measurement of the photodiode current taken over $350\,\text{s}$ at the hard X-ray branch at an energy of $1.1\,\text{keV}$. The top-up cycle of the accelerator can be seen. The mean ($\overline{x}_\text{top-up}=1.5653\times10^{-7}\,\text{A}$) and the standard deviation ($s_\text{top-up}=2.962\times10^{-10}\,\text{A}$) used for the transparency calculations are shown.}
\label{SiN:Data_filling}
\end{figure}

A longer calibration measurement, as shown in figure~\ref{SiN:Data_filling}, reveals the top-up pattern of the synchrotron as described in subsection~\ref{ssec:Soleil}. To avoid jumps in the data taken the measurements were only taken between two top-ups. However, depending on the starting point on the slope, the measurements are slightly offset from one another. Since no time-stamped data of the top-up cycle is available, the error is taken into account as a systematic error of $\epsilon_\text{top-up}= 0.2\,\%$.

At energies above $\sim 6\,\text{keV}$, a constant slope of the flux of unknown origin could be detected. Thermal behaviour of the beamline could be an explanation, as well as a problem with a calibrated diode or the ampere meter. The error here could be calculated from the means of the two measurements of the nominal flux at each energy.\\

An additional systematic uncertainty comes from the harmonics (less than 1\,\%~\cite{menesguen2011characterization}) of the crystal monochromator as described in section~\ref{ssec:Soleil}.

\subsection{Results}
\begin{figure}[htb] 
\centering 
\begin{tikzpicture} 
    \begin{axis}[
                grid = major,
                xlabel={Energy [eV]},
                ylabel=Transparency,
				xmin=10, xmax=15050,
				ymin = 0, ymax = 1.1,
				scaled x ticks=false,
				width=0.8\linewidth,
				height=0.5\linewidth,
				legend style={at={(0.97,0.03)},anchor=south east,cells={align=left}}
                ]
        \addplot[black, line width=0.5pt] table[skip first n=1, x index=0, y expr = \thisrowno{1}*\thisrowno{2}, col sep=space] {data/SiN_window_trans.csv};
		\addlegendentry{$200\,\text{nm}$ $\text{SiN}_\text{x}$ + $20\,\text{nm}$ Al,, $\rho=3.44\,\text{g}/\text{cm}^3$}
		
		\addplot[black, line width=0.5pt, dotted] table[skip first n=1, x index=0, y expr = \thisrowno{1}*\thisrowno{10}, col sep=space] {data/SiN_window_trans.csv};
		\addlegendentry{$200\,\text{nm}$ $\text{SiN}_\text{x}$ + $20\,\text{nm}$ Al, $\rho=3.0\,\text{g}/\text{cm}^3$}
				
		\addplot[blue, only marks, mark size=1pt, error bars/.cd, y dir=both, y explicit] table[skip first n=1, x index=0, y index=1, y error plus expr= sqrt((\thisrowno{2}*\thisrowno{2})+(0.002*0.002)), y error minus expr = (sqrt((\thisrowno{2}*\thisrowno{2})+(0.002*0.002))+\thisrowno{10}), col sep=space] {data/window_scan_transparencys.txt};
		\addlegendentry{$200\,\text{nm}$ window data}
				
		\addplot[red, line width=0.5pt] table[skip first n=1, x index=0, y expr = \thisrowno{1}*\thisrowno{3}, col sep=space] {data/SiN_window_trans.csv};
		\addlegendentry{$300\,\text{nm}$ $\text{SiN}_\text{x}$ + $20\,\text{nm}$ Al, $\rho=3.44\,\text{g}/\text{cm}^3$}
				
		\addplot[red, line width=0.5pt, dotted] table[skip first n=1, x index=0, y expr = \thisrowno{1}*\thisrowno{8}, col sep=space] {data/SiN_window_trans.csv};
		\addlegendentry{$300\,\text{nm}$ $\text{SiN}_\text{x}$ + $20\,\text{nm}$ Al, $\rho=3.0\,\text{g}/\text{cm}^3$}
				
		\addplot[orange, only marks, mark size=1pt, error bars/.cd, y dir=both, y explicit] table[skip first n=1, x index=0, y index=3, y error plus expr= sqrt((\thisrowno{4}*\thisrowno{4})+(0.002*0.002)), y error minus expr = (sqrt((\thisrowno{4}*\thisrowno{4})+(0.002*0.002))+\thisrowno{10}), col sep=space] {data/window_scan_transparencys.txt};
		\addlegendentry{$300\,\text{nm}$ window (first batch) data}
				
		\addplot[purple, only marks, mark size=1pt, error bars/.cd, y dir=both, y explicit] table[skip first n=1, x index=0, y index=5, y error plus expr= sqrt((\thisrowno{6}*\thisrowno{6})+(0.002*0.002)), y error minus expr = (sqrt((\thisrowno{6}*\thisrowno{6})+(0.002*0.002))+\thisrowno{10}), col sep=space] {data/window_scan_transparencys.txt};
		\addlegendentry{$300\,\text{nm}$ window (last batch) data}
    \end{axis}
\end{tikzpicture}  
\caption{Transparencies of the $\text{SiN}_\text{x}$ windows for an energy range from $10\,\text{eV}$ to $15\,000\,\text{eV}$. The data and the expected transparencies are shown as lines. Expected transparencies from \cite{henke}.} 
\label{SiN:window_mess} 
\end{figure}
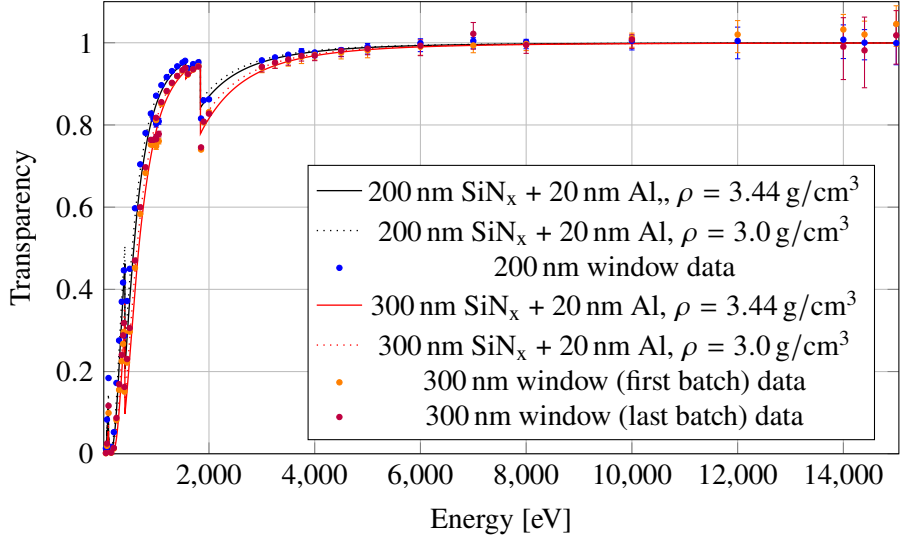

\begin{figure}[htb] 
\centering 
\begin{tikzpicture} 
    \begin{axis}[
                 grid = major,
                 xlabel={Energy [eV]},
                 ylabel=Transparency,
				xmin=10, xmax=3500,
				ymin = 0, ymax = 1.1,
				scaled x ticks=false,
				width=0.8\linewidth,
				height=0.5\linewidth,
				legend style={at={(0.97,0.03)},anchor=south east,cells={align=left}}
                ]
        \addplot[black, line width=0.5pt] table[skip first n=1, x index=0, y expr = \thisrowno{1}*\thisrowno{2}, col sep=space] {data/SiN_window_trans.csv};
		\addlegendentry{$200\,\text{nm}$ $\text{SiN}_\text{x}$ + $20\,\text{nm}$ Al, $\rho=3.44\,\text{g}/\text{cm}^3$}
				
		\addplot[black, line width=0.5pt, dotted] table[skip first n=1, x index=0, y expr = \thisrowno{1}*\thisrowno{10}, col sep=space] {data/SiN_window_trans.csv};
		\addlegendentry{$200\,\text{nm}$ $\text{SiN}_\text{x}$ + $20\,\text{nm}$ Al, $\rho=3.0\,\text{g}/\text{cm}^3$}
				
		\addplot[blue, only marks, mark size=1pt, error bars/.cd, y dir=both, y explicit] table[skip first n=1, x index=0, y index=1 ,y error plus expr= sqrt((\thisrowno{2}*\thisrowno{2})+(0.002*0.002)),y error minus expr = (sqrt((\thisrowno{2}*\thisrowno{2})+(0.002*0.002))+\thisrowno{10}) ,col sep=space] {data/window_scan_transparencys.txt};
		\addlegendentry{$200\,\text{nm}$ data}
				
		\addplot[red, line width=0.5pt] table[skip first n=1, x index=0, y expr = \thisrowno{1}*\thisrowno{3}, col sep=space] {data/SiN_window_trans.csv};
		\addlegendentry{$300\,\text{nm}$ $\text{SiN}_\text{x}$ + $20\,\text{nm}$ Al, $\rho=3.44\,\text{g}/\text{cm}^3$}
				
		\addplot[red, line width=0.5pt, dotted] table[skip first n=1, x index=0, y expr = \thisrowno{1}*\thisrowno{8}, col sep=space] {data/SiN_window_trans.csv};
		\addlegendentry{$300\,\text{nm}$ $\text{SiN}_\text{x}$ + $20\,\text{nm}$ Al, $\rho=3.0\,\text{g}/\text{cm}^3$}
				
		\addplot[orange, only marks, mark size=1pt, error bars/.cd, y dir=both, y explicit] table[skip first n=1, x index=0, y index=3, y error plus expr= sqrt((\thisrowno{4}*\thisrowno{4})+(0.002*0.002)), y error minus expr = (sqrt((\thisrowno{4}*\thisrowno{4})+(0.002*0.002))+\thisrowno{10}), col sep=space] {data/window_scan_transparencys.txt};
		\addlegendentry{$300\,\text{nm}$ first batch data}
				
		\addplot[purple, only marks, mark size=1pt, error bars/.cd, y dir=both, y explicit] table[skip first n=1, x index=0, y index=5, y error plus expr= sqrt((\thisrowno{6}*\thisrowno{6})+(0.002*0.002)), y error minus expr = (sqrt((\thisrowno{6}*\thisrowno{6})+(0.002*0.002))+\thisrowno{10}), col sep=space] {data/window_scan_transparencys.txt};
		\addlegendentry{$300\,\text{nm}$ last batch data}
				
		\addplot[blue, only marks, mark=x, mark size=4pt] table[skip first n=1, x index=0, y index=1] {data/data_SiN_off.csv};
		\addplot[orange, only marks, mark=x, mark size=4pt] table[skip first n=1, x index=0, y index=3] {data/data_SiN_off.csv};
		\addplot[purple, only marks, mark=x, mark size=4pt] table[skip first n=1, x index=0, y index=5] {data/data_SiN_off.csv};
    \end{axis}
\end{tikzpicture}  
\caption{Transparencies of the $\text{SiN}_\text{x}$ window over an energy range from $10\,\text{eV}$ to $3500\,\text{eV}$. It is clearly visible that the data points are always above the expectation (solid line). Therefore, the density of the silicon nitride membranes is supposed to be different (dashed line). Non-conclusive data points ($950\,\text{eV}$, $1000\,\text{eV}$ $1050\,\text{eV}$) are marked with crosses. Expected transparencies from \cite{henke}.} 
\label{SiN:window_mess_3500} 
\end{figure}
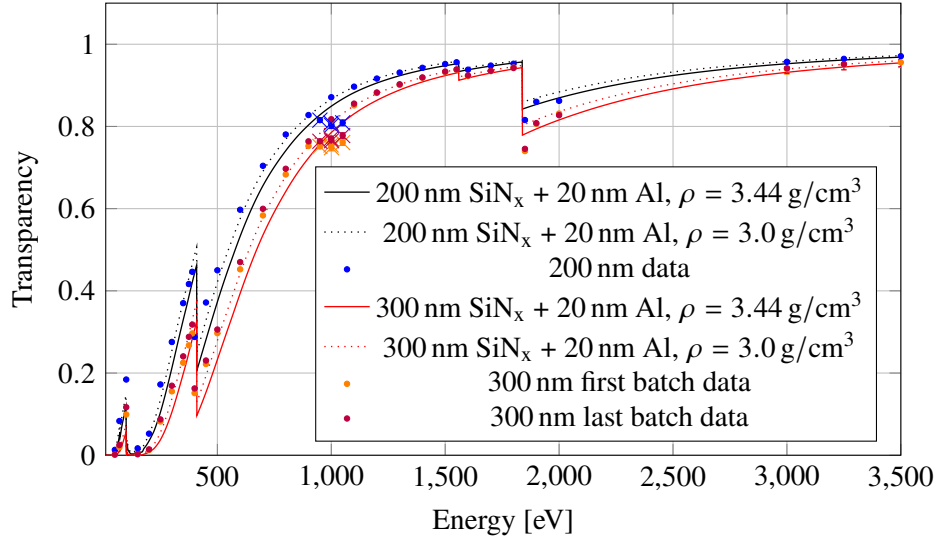

The data taken during the measurement campaign was analysed and compared to the expected transparency curves. The systematic errors described in section~\ref{ssec:Datataking} are included. With the means of the diode (nominal flux) $\overline{x}_{\text{diode}, n}$ and the window $\overline{x}_{i, n}$ currents ($i$ denoting the window type and $n$ denoting the energy), the transparencies can be computed
\begin{align}
	t_{i, n} &= \frac{\overline{x}_{i, n}}{\overline{x}_{\text{diode}, n}}\, .
\end{align}
 
The results are shown in figure~\ref{SiN:window_mess}. The overall picture shows that the data fits with the simulated transparencies. However, at low energies, as shown in figure~\ref{SiN:window_mess_3500}, the measurements do not fully represent the expectation. Therefore, either the thickness of the membranes is thinner than expected, or the composition and density are different. Since the manufacturing tolerances are too small to explain the behavior, the latter is the more likely explanation. Silicon-rich silicon nitride thin films can be produced in a huge variety of compositions and densities, which is why no real selection can be made. However, for a slightly reduced density, from $\rho_\text{nom}=3.44\,\text{g}/\text{cm}^3$ to $\rho_\text{new}=3.0\,\text{g}/\text{cm}^3$ the data agrees much better with the expectation. One setting at the low-energy branch showed non-conclusive results. The data points from those measurements are marked and should not be taken into account. 

\section{Summary}
We developed, together with the company Norcada, large diameter ($14\,\text{mm}$) highly transparent X-ray windows. This gives access to sub-keV X-rays for multiple purposes, where the mode of operation is outside of a vacuum. We showed the durability and stability at pressure differences of $1.5\,\text{bar}$. It could also be shown that the windows are vacuum tight (leak rate smaller than $3.0\times10^{-9}\,\text{mbar\,l\,s}^{-1}$). The transparencies of the windows were measured from $50\,\text{eV}$ to $15\,\text{keV}$, showing better than expected results. At 1\,keV transparencies as high as 80\,\% were achieved.

\acknowledgments
The authors acknowledge support from the Agence Nationale de la Recherche (France)ANR-19-CE31-0024 and from the BMBF (Germany) 05H21PDRD2. The authors acknowledge SOLEIL for provision of synchrotron radiation facilities and we would like to thank Pascal Mercere and Paulo Da Silva for assistance in using  METROLOGIE beamline. We acknowledge the support and work from the company NORCADA for developing and providing the windows.

% Bibliography

%% [A] Recommended: using JHEP.bst file
\bibliographystyle{JHEP}
\bibliography{references.bib}

%% or
%% [B] Manual formatting (see below)
%% (i) We suggest to always provide author, title and journal data or doi:
%% in short all the informations that clearly identify a document.
%% (ii) please avoid comments such as "For a review'', "For some examples",
%% "and references therein" or move them in the text. In general, please leave only references in the bibliography and move all
%% accessory text in footnotes.
%% (iii) Also, please have only one work for each \bibitem.

\end{document}